\documentclass[showpacs, amsmath,amsfonts,amssymb, aps]{revtex4}

\usepackage{tabularx}
\usepackage{enumitem}
\usepackage{graphicx}
\usepackage{float}
\usepackage{dcolumn}
\usepackage{bm}
\usepackage{mathrsfs}
\usepackage{amsmath}

\usepackage{amsfonts}

\begin{document}

\title{A comment on singular and non-singular black holes using the Gaussian distribution}

\author{D. Batic}
\email{davide.batic@ku.ac.ae}
\affiliation{%
Department of Mathematics,\\  Khalifa University of Science and Technology,\\ Main Campus, Abu Dhabi,\\ United Arab Emirates}
\author{M. Nowakowski}
\email{marek.nowakowski@ictp-saifr.org}
\affiliation{
ICTP South American Institute for Fundamental Research (ICTP-SAIFR)\\
IFT-UNESP, R. Bento Teobaldo Ferraz 271, Bloco 2\\
01140-070 Sao Paulo, SP, Brazil
}
\author{S. A. Salim}
\email{100053894@ku.ac.ae}
\affiliation{
Department of Mathematics,\\  Khalifa University of Science and Technology,\\ Main Campus, Abu Dhabi,\\ United Arab Emirates}

\date{\today}

\begin{abstract}
In this work, we join the controversial discussion on singular and non-singular black holes using the Gaussian distribution. Our result which uses correct boundary conditions shifts the debate in favour of regular black holes at the centre. The present findings add new insights into the ongoing discussions surrounding singularities in black hole solutions of the Einstein equations.
    
\end{abstract}
\maketitle

\section{Introduction}
Black holes are fascinating objects predicted by Einstein's theory of general relativity. They are characterized by event horizons, which mark a point of no return beyond which not even light can escape the black hole's strong gravitational pull. The study of black holes has led to many breakthroughs in fundamental physics, such as the discovery of Hawking radiation, a process by which black holes evaporate and lose mass. One of the key features of black holes is the presence of a singularity at the center of the black hole, where the curvature of space-time becomes infinite. It is a point beyond which our current understanding of physics breaks down, and it is often seen as a limitation of the theory of general relativity. In recent years, there has been growing interest in the study of regular black holes, which are solutions to Einstein's equations that do not contain singularities \cite{Bardeen,Nicolini0,BN,Modesto,Zhang,Malafarina,Haggard,Rov}. These regular black holes are characterized by a non-zero size at their centers, and are expected to exhibit physical properties different from those of traditional black holes. The search for regular black hole solutions is motivated by both theoretical and observational considerations. On the one hand, regular black holes could provide new insights into the behavior of space-time at extreme scales and the nature of gravity. On the other hand, regular black holes could also have astrophysical implications, as they may play a role in the formation and evolution of galaxies and the universe as a whole. In this context, the aim of this paper is to provide solid evidence disproving the arguments presented by \cite{Chinaglia} against the regular black hole solutions. 

In 2006, \cite{Nicolini} showed how coordinate noncommutativity can cure the central singularity of a Schwarzschild black hole. In particular, by considering a mass $M$ diffused over a given region according to a Gaussian distribution and by modelling the matter content in terms of an anisotropic fluid subject to the equation of state $p_r=-\rho$ the authors in \cite{Nicolini} succeeded in obtaining a new solution to the Einstein field equations where the central singularity is replaced by a regular de Sitter core. The aforementioned paper is focused on a specific type of black hole solutions known as "regular black holes". These black holes are solutions of Einstein's field equations in which the singularity at the center of the black hole is replaced by a regular core, where the physical properties are well-defined and the curvature remains finite. The author argued that regular black holes could provide a more physical and consistent description of the gravitational collapse, avoiding the problems associated with the presence of singularities. \cite{Nicolini} also provided a possible explanation of the mechanism behind the formation of regular black holes, which could be relevant in the study of the quantum nature of gravity. 

On the other hand, \cite{Chinaglia} in "A note on singular and non-singular black holes" argue that the curvature singularity is a fundamental feature of the solutions of the Einstein equations and cannot be removed by adding new terms to the action, as is commonly done in models of regular black holes. We briefly mention that additional terms in the action are considered in order to incorporate effects due to vacuum polarization and particle creation in the interior of a black hole. Other approaches leading to regular black hole solutions follow a different path where one considers modifications of gravity such as in the $f(R)$ gravity, Gauss-Bonnet theory or introduces non local modifications of general relativity such ghost free gravity. For an overview of regular black hole solutions we refer to \cite{Chinaglia,Nicolini3,Ansoldi,Frolov}. The authors in \cite{Chinaglia} also point out that, even in the presence of these new terms, the metric remains singular at the center of the black hole, which they claim is indicative of the presence of a curvature singularity. However, this claim has been the subject of much debate and further research is needed to fully understand the nature of the singularities in these solutions. 

In view of the result presented in the next section, it is appropriate to present here the original argument proposed by \cite{Chinaglia} according to which the noncommutative geometry inspired Schwarzschild solution should not be regular at $r=0$. Assuming a spherically symmetric, static and asymptotically flat manifold with the line element
\begin{equation}\label{le}
ds^2=g_{\mu\nu}dx^\mu dx^\nu,\quad
(g_{\mu\nu})=\mbox{diag}(-f(r),1/f(r),r^2,r^2\sin^2{\theta}),
\end{equation}
the Einstein field equations in geometrized units
\begin{equation}
G_{\mu\nu}=-8\pi T_{\mu\nu},\quad 
(T^\mu{}_\nu)=\mbox{diag}(\rho,p_r,p_\bot,p_\bot),\quad p_r=-\rho
\end{equation}
boil down to the following non-homogeneous first order differential equation
\begin{equation}\label{ODE}
r\frac{df}{dr}+f(r)-1=-8\pi r^2\rho.
\end{equation}
At this point, they find the general solution of (\ref{ODE}), i.e.
\begin{equation}
f(r)=1-\frac{c}{r}-\frac{8\pi}{r}\int_0^r du~u^2\rho(u),
\end{equation}
where $c$ is an arbitrary integration constant and they conclude that independently of the regularity of the energy density $\rho$ the metric coefficient $f(r)$ is always plagued by a curvature singularity at $r=0$. According to them even in the absence of matter, i.e. $\rho=0$, the corresponding solution which in this case is the Minkowski metric in spherical coordinates will be singular at $r=0$! Hence, their conclusion is that \cite{Nicolini3} derived the modified Schwarzschild solution by assuming $c=0$ without providing any justification. Recently, \cite{Nicolini3} offered several physical arguments against the conclusions in \cite{Chinaglia}. First of all, they reiterated the famous Feynman mantra according to which extreme care must be exercised in order to distinguish between vacuum solutions and solutions in vacuum \cite{Feynman}. Moreover, two counter-arguments involving a charged shell and the de Sitter metric were brought forward by \cite{Nicolini3} where they showed that following the line of reasoning in \cite{Chinaglia} one should conclude that both the electric potential inside a charged shell and the de Sitter metric would be singular at $r=0$!

Interestingly, both \cite{Chinaglia} and \cite{Nicolini3} missed an important observation made by \cite{Wald} (see page $126$) which is worth to be summarised here below because it  points against the conclusions of \cite{Chinaglia}. First of all, since the line element (\ref{le}) is static, it is also stationary, and therefore, there must exist a one-parameter group of isometries whose orbits are represented by time-like curves. This can also be paraphrased in an equivalent form by saying that a stationary Lorentzian manifold always exhibits a time-like Killing vector field. Moreover, being the space-time static,  it must have a space-like hypersurface $\Sigma$ which is orthogonal to the orbits of the isometry. \cite{Wald} goes on by solving (\ref{ODE}) which is represented therein in the following equivalent form
\begin{equation}\label{ef}
\frac{1}{r^2}\frac{d}{dr}\left[r(f(r)-1)\right]=-8\pi\rho
\end{equation}
for the case of a perfect fluid. Even though \cite{Wald} ends up with the same general solution as \cite{Chinaglia}, i.e.
\begin{equation}
f(r)=1-\frac{2m(r)}{r},\quad m(r)=4\pi\int_0^r du~u^2\rho(u)+C
\end{equation}
with $C=c/2$, he makes the crucial observation that requirement of the metric being smooth on $\Sigma$ at $r=0$ requires that $f(r)\to 1$ as $r\to 0$. According to this reasoning, \cite{Wald} sets $C=0$. At this point, we would also like to add that recasting (\ref{ODE}) into the form (\ref{ef}) is quite enlightening because it tells us that if one starts with an energy density $\rho$ which is smooth for all $r\geq 0$ like it is indeed the case for the Gaussian distribution used by \cite{Nicolini3},  the left hand side (l.h.s) of (\ref{ef}) must also be a smooth function. On the other hand, if one keeps insisting as in \cite{Chinaglia} that $f(r)$ must contain the term $-c/r$, then its substitution into the l.h.s. of (\ref{ef}) results into the production of the term $-1/r^2$ which in turn implies that $\rho$ is not smooth at $r=0$. This is clearly a contradiction and we conclude that the argument of \cite{Chinaglia} for the presence of a curvature singularity in regular black holes is mathematically flawed. In the next section, we will present a second argument in support of the regularity of the Noncommutative Geometry inspired Schwarzschild black hole.

\section{The Newtonian gravitational potential of a smeared particle}
We consider the problem of finding the Newtonian gravitational potential in the case of particle whose mass density is described by a Gaussian distribution \cite{Nicolini3}, namely
\begin{equation}\label{density}
\rho_{\theta}(r)=M\Xi_\theta(r),\quad
\Xi_\theta(r)=\frac{e^{-r^2/(4\theta)}}{(4\pi\theta)^{3/2}}.
\end{equation}
In particular, we show that the resulting Newtonian potential $\Phi(r)$ is regular at $r=0$ and  explain the reason why the $g_{00}$ metric coefficient of the Noncommutative Geometry inspired Schwarzschild solution must also be regular at the origin. Note that by considering (\ref{density}) we replaced the usual point-like particle assumption with a smeared object whose mass follows a Gauss distribution of standard deviation $\sigma=\sqrt{2\theta}$. In the context of Noncommutative Geometry, $\theta$ is the so-called noncommutative parameter and it has the dimensions of length. As previously noted in \cite{Nicolini0}, Noncommutativity Geometry is anticipated to have significance on the scale range of $\ell_p<\sqrt{\theta}<10^{-16}$ m, where the Planck length $\ell_p$ is $1.6\times 10^{-35}$ m. Since this scale is significantly small, the mass distribution of a smeared object is usually characterized by a Gaussian with a very narrow peak.

Due to the spherical symmetry of (\ref{density}), the Newtonian potential per unit mass $\widehat{\Phi}=\Phi/M$ associated to a fuzzy particle is obtained by solving the Poisson equation in natural units ($G_N=1$)
\begin{equation}\label{ODEg}
\frac{1}{r^2}\frac{d}{dr}\left(r^2\frac{d\Phi}{dr}\right)=4\pi \Xi_\theta(r)
\end{equation}
subject to the constraints
\begin{enumerate}
\item
$\widehat{\Phi}(r)\to 0$ as $r\to+\infty$; 
\item
$\widehat{\Phi}(r)\to -1/r$ in the limit of $\theta\to 0^+$.
\end{enumerate}
The second constraint is due to the fact that the sequence of Gaussians $(\rho_\theta)$ converges to a Dirac delta. For a rigorous proof we refer to Appendix~\ref{A1}. The general solution to (\ref{ODEg}) reads
\begin{equation}\label{gensol}
\widehat{\Phi}(r)=c_1+\frac{c_2}{r}-\frac{1}{r}\mbox{erf}\left(\frac{r}{2\sqrt{\theta}}\right)
\end{equation}
with $c_1$ and $c_2$ arbitrary integration constants. Here $\mbox{erf}$ denotes the error function. Since $\mbox{erf}(r/2\sqrt{\theta})\to 1$ as $r\to\infty$ \cite{Abra}, the condition $\widehat{\Phi}(r)\to 0$ as $r\to+\infty$ requires that $c_1=0$. Regarding the second integration constant, we observe that as $\theta\to 0^{+}$ the function $\widehat{\Phi}(r)$ must reproduce the usual Newtonian potential associated to a point-like particle whose mass density is represented by a Dirac delta distribution, that is
\begin{equation}
\lim_{\theta\to 0^+}\widehat{\Phi}(r)=-\frac{1}{r}.
\end{equation}
For $\theta\to 0^+$ and fixed $r$ we can consider the following asymptotic expansion for the error function \cite{Abra}
\begin{equation}
\mbox{erf}\left(\frac{r}{2\sqrt{\theta}}\right)=1-\frac{2}{\sqrt{\pi}r}\theta^{1/2}e^{-\frac{r^2}{4\theta}}+\mathcal{O}\left(\theta^{3/2}e^{-\frac{r^2}{4\theta}}\right).
\end{equation}
A trivial application of the squeeze theorem shows that both terms $\theta^{1/2}e^{-\frac{r^2}{4\theta}}$ and $\theta^{3/2}e^{-\frac{r^2}{4\theta}}$ converges to zero as $\theta\to 0^+$. This implies that pointwise in $r$ we have $\mbox{erf}(r/2\sqrt{\theta})\to 1$ as $\theta\to 0^+$. If we take the limit $\theta\to 0^+$ on both sides of (\ref{gensol}), we find that
\begin{equation}
-\frac{1}{r}=\frac{c_2}{r}-\frac{1}{r}
\end{equation}
from which it follows that $c_2=0$. Hence, the Newtonian potential for a smeared particle is given in SI units by
\begin{equation}\label{GPSP}
\Phi(r)=-\frac{G_N M}{r}\mbox{erf}\left(\frac{r}{2\sqrt{\theta}}\right).
\end{equation}
Furthermore, using the expansion \cite{Abra}
\begin{equation}
\mbox{erf}\left(\frac{r}{2\sqrt{\theta}}\right)=\frac{r}{\sqrt{\pi\theta}}+\mathcal{O}(r^2),
\end{equation}
one immediately realize that (\ref{GPSP}) is regular at $r=0$ where it takes a finite value. This can be easily seen by  expanding the error function around $r=0$ \cite{Abra} as follows
\begin{equation}
\lim_{r\to 0^+}\Phi(r)=-G_N M\lim_{r\to 0^+}\frac{1}{r}\left[\frac{r}{\sqrt{\pi\theta}}+\mathcal{O}(r^3)\right]=-\frac{G_N M}{\sqrt{\pi\theta}}.
\end{equation}
At this point a remark is in order. It is well-known that in the Newtonian limit the metric tensor can be approximated by $g_{ab}=\eta_{ab}+h_{ab}$ where $\eta_{ab}$ is the Minkowski metric tensor, $h_{ab}$ is a small correction and 
\begin{equation}
g_{00}=1+2\Phi(r),
\end{equation}
where $\Phi(r)$ satisfies the Poisson equation. According to the discussion above the corresponding $g_{00}$ is regular at $r=0$. Moreover, for $\theta\to 0^+$ we have $g_{00}\to 1-\frac{2M}{r}$ as one would expect in the case of the classic Schwarzschild solution. This signalizes that even without knowing the exact expression of the noncommutative geometry inspired solution obtained in \cite{Nicolini}, its Newtonian limit exhibits the property of being regular at $r=0$. Our reasoning shows that if we work within a static spherically symmetric geometry and within a regular matter distribution, the corresponding Newtonian limit is regular at the origin. The argument presented in this section, in addition to the one discussed in the previous section, refutes the claim made by \cite{Chinaglia} regarding regular black hole solutions. Thus, we agree with \cite{Nicolini3} that the conclusions reached by \cite{Chinaglia} are invalid.

At the end, we notice that the regular black holes are not limited to mini black holes, but could have applications to supermassive black holes at the centre of galaxies in connection with Dark Matter \cite{D1,D2}.

\section{Conclusions}

It is important to recognize that scientific theories and conclusions are not necessarily proven or disproven by a single paper or author. The scientific process involves continuous testing and evaluation of evidence, which may result in the revision of individual studies or challenge existing ideas with some gaps of time between different results. In this study, we have demonstrated that black holes with a de Sitter equation of state and Gaussian mass distribution have indeed a regular centre. Our findings align with the views of \cite{Nicolini3} and add new insights to the ongoing debate about the nature of singularities in black hole solutions of the Einstein field equations. By contributing to this debate, we hope to promote further understanding and development in this field.

\appendix
\section{The convergence analysis of the sequence of Gaussian distributions $(\rho_\theta)$}\label{A1}
For $\rho_\theta$ given as in (\ref{density}) we study the convergence problem of the sequence $(\rho_\theta)$ in the limit $\theta\to 0^+$. In order for the sequence of functions $(\rho_\theta)$ to define a Dirac delta distribution, we must have 
\begin{enumerate}
\item
$\int_{\mathbb{R}^3}\rho_\theta(r)d^3{\bf{x}}=M$;
\item
For every smooth function $f$ with compact support $K=[0,a]$ on the real line 
\begin{equation}\label{TBS}
\lim_{\theta\to 0^+}\int_{\mathbb{R}^3}\rho_\theta(r)f(r)d^3{\bf{x}}=Mf(0).
\end{equation}
If this is the case, then we write
\begin{equation}
\lim_{\theta\to 0^+}\rho_\theta(r)=M\delta(r).
\end{equation}
\end{enumerate}
Going to spherical coordinates with $d^3{\bf{x}}=r^2\sin{\vartheta}dr d\vartheta d\varphi$ and taking into account that
\begin{equation}
\int_0^\infty r^2 e^{-r^2/(4\theta)}dr=\sqrt{4\pi}\theta^{3/2},
\end{equation}
the check of the first condition reduces to a trivial computation. To verify condition 2., we observe that proving (\ref{TBS}) is equivalent to show that
\begin{equation}\label{TBS1}
\lim_{\theta\to 0^+}\int_{0}^\infty r^2\rho_\theta(r)\left[f(r)-f(0)\right]dr=0.
\end{equation}
Note that (\ref{TBS1}) is obtained from (\ref{TBS}) by first expressing $M$ as the  volume integral appearing in condition 1. followed by integration over the angular variables. Using the fact that $f$ has compact support and applying the triangle inequality yields
\begin{equation}\label{TE}
\left|\int_{0}^\infty r^2\rho_\theta(r)\left[f(r)-f(0)\right]dr\right|\leq
\left|\int_{0}^a r^2\rho_\theta(r)\left[f(r)-f(0)\right]dr\right|+|f(0)|\int_{a}^\infty r^2\rho_\theta(r)~dr.
\end{equation}
A straightforward computation shows that
\begin{eqnarray}
\int_{a}^\infty r^2\rho_\theta(r)~dr&=&\frac{2M}{(4\pi)^{3/2}}\left[-\frac{1}{\sqrt{\theta}}\lim_{r\to+\infty}re^{-\frac{r^2}{4\theta}}+\sqrt{\pi}\lim_{r\to+\infty}\mbox{erf}\left(\frac{r}{2\sqrt{\theta}}\right)+\frac{a}{\sqrt{\theta}}e^{-\frac{a^2}{4\theta}}-\sqrt{\pi}\mbox{erf}\left(\frac{a}{2\sqrt{\theta}}\right)\right],\\
&=&\frac{2M}{(4\pi)^{3/2}}\left[\sqrt{\pi}-\sqrt{\pi}\mbox{erf}\left(\frac{a}{2\sqrt{\theta}}\right)+\frac{a}{\sqrt{\theta}}e^{-\frac{a^2}{4\theta}}\right],\label{A7}
\end{eqnarray}
where we used the property that $\mbox{erf}(x)\to 1$ as $x\to +\infty$. Finally, letting $\theta\to 0^+$ in (\ref{A7}) gives
\begin{equation}
\lim_{\theta\to 0^+}\int_{a}^\infty r^2\rho_\theta(r)~dr=0
\end{equation}
and in the same limit, (\ref{TE}) becomes
\begin{equation}
\lim_{\theta\to 0^+}\left|\int_{0}^\infty r^2\rho_\theta(r)\left[f(r)-f(0)\right]dr\right|\leq\lim_{\theta\to 0^+}\left|\int_{0}^a r^2\rho_\theta(r)\left[f(r)-f(0)\right]dr\right|.
\end{equation}
Since $f$ is continuous on $[0,r]$ with $r\leq a$ and differentiable on $(0,r)$, then by the Mean Value Theorem there exists a $c\in(0,r)$ such that $f(r)-f(0)=f^{'}(c)r$. Let $\Gamma=\max_{r\in K}{\{|f^{'}(c)|\}}$ and observe that $|f(r)-f(0)|\leq\Gamma r$. This implies that
\begin{eqnarray}
\lim_{\theta\to 0^+}\left|\int_{0}^\infty r^2\rho_\theta(r)\left[f(r)-f(0)\right]dr\right|&\leq&\frac{\Gamma M}{(4\pi)^{3/2}}\lim_{\theta\to 0^+}\frac{1}{\theta^{3/2}}\int_{0}^a r^3 e^{-\frac{r^2}{4\theta}}dr,\\
&=&\frac{2\Gamma M}{(4\pi)^{3/2}}\lim_{\theta\to 0^+}\left[4\theta^{1/2}-\frac{a^2}{\theta^{1/2}}e^{-\frac{a^2}{4\theta}}-4\theta^{1/2}e^{-\frac{a^2}{4\theta}}\right]=0
\end{eqnarray}
and the proof is completed.

{\bf{Data accessibility}} This article does not use data.

\section*{Author contributions statement}
D.B. conceived the problem discussed in this work and contributed to the interpretation of the results. M.N. participated in the preparation of section I.   D.B. and S.S. worked out all of the technical details and performed the  calculations. D.B. wrote the manuscript in consultation with M.N.. All authors reviewed the manuscript.


\begin{thebibliography}{99}
\bibitem{Bardeen}
J.M. Bardeen, {\it{Non-singular general-relativistic gravitational collapse}}, Proceedings of the International Conference GR5, Tbilisi, U.S.S.R., p. 174 (1968).
\bibitem{Nicolini0}
P. Nicolini, {\it{Noncommutative black holes, the final appeal to quantum gravity: a review}},  Int. J. Mod. Phys. A {\bf{24}}, 1229 (2009).
\bibitem{BN}
I. Arraut, D. Batic, M. Nowakowski, {\it{A noncommutative model for a mini black hole}}, Class. Quantum Gravity {\bf{26}}, 245006 (2009).
\bibitem{Modesto}
L. Modesto, {\it{Space-Time Structure of Loop Quantum Black Hole}}, Int. J. Theor. Phys. {\bf{49}}, 1649 (2010).
\bibitem{Zhang}
Y. Zhang, Y. Zhu, L. Modesto et C. Bambi, {\it{Can static regular black holes form from gravitational collapse?}}, Europ. Phys. J. C {\bf{75}}, 96 (2015).
\bibitem{Malafarina}
D. Malafarina and B. Toshmatov, {\it{Connection between regular black holes in nonlinear electrodynamics and semiclassical dust collapse}}, Phys. Rev. D {\bf{105}}, L121502 (2022).
\bibitem{Haggard}
H. M. Haggard and C. Rovelli, {\it{Black hole fireworks: quantum-gravity effects outside the horizon spark black to white hole tunneling}}, Phys. Rev. D {\bf{92}}, 104020 (2015).
\bibitem{Rov}
A. Barrau and C. Rovelli, {\it{Planck star phenomenology}}, Phys. Lett. B {\bf{739}}, 405 (2014).
\bibitem{Chinaglia}
S. Chinaglia and S. Zerbini, {\it{A note on singular and non-singular black holes}}, Gen. Rel. Grav. {\bf{49}}, 75 (2017).
\bibitem{Nicolini} 
P. Nicolini, A. Smailagic and E. Spallucci, {\it Noncommutative geometry inspired Schwarzschild black hole}, Phys. Lett. B {\bf{632}}, 547 (2006).
\bibitem{Nicolini3}
P. Nicolini, A. Smailagic and E. Spallucci, {\it{Remarks on regular black holes}}, Int. J. Geom. Meth. Mod. Phys. {\bf{15}}, 1850018 (2018).
\bibitem{Ansoldi}
S. Ansoldi, {\it{Spherical black holes with regular center: a review of existing models including a recent realization with Gaussian sources}}, KUNS-2108, arXiv:0802.0330 [gr-qc] (2008).
\bibitem{Frolov}
V. P. Frolov, {\it{Remarks on non-singular black holes}}, EPJ Web of Conferences {\bf{168}}, 01001 (2018).
\bibitem{Feynman}
R. P. Feynman, R. B. Leighton and M. Sands, {\it{The Feynman Lectures on Physics}}, Vol. II: {\it{Mainly Electromagnetism and Matter}}, Basic Books; New Millennium ed. (2011).
\bibitem{Wald}
R. M. Wald, {\it{General Relativity}}, The University Chicago Press, Chicago and London (1984).
\bibitem{Abra}
M. Abramowitz and I. A. Stegun, {\it{Handbook of Mathematical Functions: With Formulas, Graphs, and Mathematical Tables}}, Dover Publications, New York (1965).
\bibitem{D1}
D. Batic, D. A. Abuhejleh and M. Nowakowski, {\it{Fuzzy dark matter black holes and droplets}}, Eur. Phys. J. C {\bf{81}}, 777 (2021).
\bibitem{D2}
D. Batic, J. M. Faraji and M. Nowakowski, {\it{Possible Connection between Dark Matter and Supermassive Black Holes}}, Eur. Phys. J. C {\bf{82}}, 759 (2022).
\end{thebibliography}
\end{document}